\documentstyle[12pt,axodraw]{article}

\catcode`@=12
\def\gsim{\:\raisebox{-0.5ex}{$\stackrel{\textstyle>}{\sim}$}\:} 
\begin{document}
\thispagestyle{empty}
\begin{flushright} 
UCRHEP-T320\\ 
TIFR/TH01-39 \\
October 2001\
\end{flushright}
\vspace{0.5in}
\begin{center}
{\Large \bf Gauged L$_\mu$--L$_\tau$ with Large Muon Anomalous Magnetic\\ 
Moment and the Bimaximal Mixing of Neutrinos\\}
\vspace{1.5in}
{\bf Ernest Ma$^a$, D. P. Roy$^b$, and Sourov Roy$^b$\\}
\vspace{0.2in}
{\sl $^a$ Physics Department, University of California, Riverside, 
California 92521\\}
\vspace{0.1in}
{\sl $^b$ Tata Institute of Fundamental Research, Mumbai (Bombay) 400005, 
India\\}
\vspace{1.5in}
\end{center}
\begin{abstract}\
We consider the gauging of $L_\mu - L_\tau$ as an explanation of a possibly 
large muon anomalous magnetic moment.  We then show how neutrino masses 
with bimaximal mixing may be obtained in this framework.  We study the 
novel phenomenology of the associated gauge boson in the context of present 
and future high-energy collider experiments.
\end{abstract}
\newpage
\baselineskip 24pt

In the minimal standard model of quarks and leptons with no right-handed 
neutrino singlet, one of the three lepton number differences ($L_e - L_\mu$, 
$L_e - L_\tau$, $L_\mu - L_\tau$) is anomaly-free and may be gauged 
\cite{hjlv}.  If \underline {one} right-handed neutrino singlet $N_R$ is 
added, then one of the three combinations ($B - 3L_e$, $B - 3L_\mu$, $B - 
3L_\tau$) is also anomaly-free and may be gauged \cite{mmsmr,mrsmr}.  For 
example, we could have both $L_e - L_\mu$ and $B - 3 L_\tau$.  On the other 
hand, even with just one $N_R$, we may choose to consider $L_\mu - L_\tau$ as 
the only additional gauge symmetry.

Specifically, under this extra gauge symmetry $U(1)_X$, $(\nu_\mu, \mu)_L$, 
$\mu_R$ have charge $+1$; $(\nu_\tau, \tau)_L$, $\tau_R$ have charge $-1$; 
all other fields including $N_R$ have charge $0$.  
It has already been noted \cite{bdhk} that the extra gauge boson $X$ of this 
model contributes to the muon anomalous magnetic moment as shown in Fig.~1.  
Its contribution \cite{vector} is easily calculated to be
\begin{equation}
\Delta a_\mu = {g_X^2 m_\mu^2 \over 12 \pi^2 M_X^2}.
\end{equation}
To complete the model, we add two extra Higgs doublets: $(\eta_1^+, \eta_1^0)$ 
have charge $+1$ and $(\eta_2^+, \eta_2^0)$ have charge $-1$.  This differs 
from Model C of Ref.~\cite{bdhk} in their $U(1)_X$ charge assignments.  Thus 
our model has no flavor-changing couplings in the charged-lepton sector, but 
because we also add the one $N_R$, realistic neutrino oscillations are 
allowed, as shown below.

The mass matrix spanning $X$ and the standard $Z$ boson is given by
\begin{equation}
{\cal M}^2_{XZ} = \left[ \begin{array} {c@{\quad}c} 2 g_X^2 (v_1^2 + v_2^2) 
& g_X g_Z (v_1^2 - v_2^2) \\ g_X g_Z (v_1^2 - v_2^2) & (g_Z^2/2) (v_0^2 + 
v_1^2 + v_2^2) \end{array} \right],
\end{equation}
where $v_{0,1,2}$ are the vacuum expectation values of the standard-model 
$\phi^0$ and $\eta^0_{1,2}$ respectively, with $v_0^2 + v_1^2 + v_2^2 = 
(2 \sqrt 2 G_F)^{-1}$.  If we assume $v_1 = v_2$, then there is no $X-Z$ 
mixing and $M_X = 2 g_X v_1$.  This implies
\begin{equation}
\Delta a_\mu = {m_\mu^2 \over 48 \pi^2 v_1^2} > {G_F m_\mu^2 \over 6 \pi^2 
\sqrt 2} = 1.555 \times 10^{-9}.
\end{equation}
In other words, such a model actually predicts a \underline {lower} bound on 
$\Delta a_\mu$. 

Experimentally, the muon magnetic moment has been measured precisely \cite{g-2}
and a large positive discrepancy \cite{cm} of $(4.26 \pm 1.65) \times 10^{-9}$ 
from the prediction of the standard model is possible, although there is no 
universal consensus regarding the uncertainties of the hadronic contributions 
\cite{uncertain}.  Note that $M_X/M_Z = 2 \sqrt 2 (g_X/g_Z)(v_1/\sqrt {v_0^2 
+ 2 v_1^2})$, which means that $M_X$ is allowed to be much heavier than $M_Z$ 
even though Eq.~(3) is independent of it.  For example, if $v_0 = v_1 = v_2$, 
then $\Delta a_\mu = 2.33 \times 10^{-9}$, and $M_X/g_X \simeq 200$ GeV.

To obtain a desirable pattern of neutrino masses to explain the atmospheric 
\cite{atm} and solar \cite{solar} neutrino data, we add a singlet charged 
scalar $\zeta^+$ which also has a $U(1)_X$ charge of $+1$ (but since it is 
a scalar, it does not contribute to the axial vector anomaly), and supplement 
our model with a discrete $Z_2$ symmetry, under which $\eta_{1,2}$ and $N_R$ 
are odd but all other fields are even.  The relevant Yukawa interaction terms 
are then given by
\begin{equation}
{\cal L}_Y = f_1 \bar N_R (\nu_\mu \eta_2^0 - \mu_L \eta_2^+) + f_2 \bar N_R 
(\nu_\tau \eta_1^0 - \tau_L \eta_1^+) + h \zeta^+ (\nu_e \tau_L - e_L 
\nu_\tau) + H.c.
\end{equation}
Since $N_R$ is allowed a large Majorana mass $M_N$, the canonical seesaw 
mechanism \cite{seesaw} generates one small neutrino mass
\begin{equation}
m_3 = {f_1^2 v_2^2 + f_2^2 v_1^2 \over M_N} = {2 f_1^2 v_1^2 \over M_N}
\end{equation}
corresponding to the eigenstate
\begin{equation}
\nu_3 = {f_1 \nu_\mu + f_2 \nu_\tau \over \sqrt {f_1^2 + f_2^2}}.
\end{equation}
We now allow the $Z_2$ discrete symmetry to be broken \underline {softly}, 
i.e. by terms of dimension 2 or 3 in the Lagrangian.  However, given the 
gauge symmetry and particle content of our model, the only possible such 
term is the trilinear scalar interaction
\begin{equation}
{\cal L}_S =  \lambda ~\zeta^- (\eta_1^+ \phi^0 - \eta_1^0 \phi^+) + H.c.
\end{equation}
This generates a radiative $\nu_e \nu_\tau$ mass as shown in Fig.~2.  As a 
result, the $3 \times 3$ neutrino mass matrix in the $(\nu_e, \nu_\mu, 
\nu_\tau)$ basis is given by
\begin{equation}
{\cal M}_\nu = \left[ \begin{array} {c@{\quad}c@{\quad}c} 0 & 0 & m' \\ 
0 & m_3 c^2 & m_3 s c \\ m' & m_3 s c & m_3 s^2 \end{array} \right],
\end{equation}
where $s \equiv \sin \theta$ and $c \equiv \cos \theta$ with $s/c = f_2/f_1$. 
Assuming $m'$ to be much smaller than $m_3$, the eigenvalues are easily 
determined to be
\begin{equation}
\pm c m' - {s^2 m'^2 \over 2 m_3}, ~~~ m_3 + {s^2 m'^2 \over m_3},
\end{equation}
corresponding to the eigenstates
\begin{eqnarray}
\nu_1 &=& {1 \over \sqrt 2} \left[ 1-{s^2 m' \over 4 c m_3} \right] \nu_e - 
{s \over \sqrt 2} \left[ 1 + {(4-3s^2)m' \over 4 c m_3} \right] \nu_\mu + 
{c \over \sqrt 2} \left[ 1 - {3 s^2 m' \over 4 c m_3} \right] \nu_\tau, \\ 
\nu_2 &=& {1 \over \sqrt 2} \left[ 1 + {s^2 m' \over 4 c m_3} \right] \nu_e + 
{s \over \sqrt 2} \left[ 1 - {(4-3s^2) m' \over 4 c m_3} \right] \nu_\mu - 
{c \over \sqrt 2} \left[ 1 + {3s^2 m' \over 4 c m_3} \right] \nu_\tau, \\ 
\nu_3 &=& {s m' \over m_3} \nu_e + c \nu_\mu + s \nu_\tau.
\end{eqnarray}
If $f_1 = f_2$ so that $s = c = 1/\sqrt 2$, we then obtain nearly bimaximal 
mixing of neutrinos for understanding the atmospheric and solar data as 
neutrino oscillations.  In addition,
\begin{eqnarray}
\Delta m_{23}^2 \simeq \Delta m_{13}^2 &\simeq& m_3^2 + (2s^2-c^2)m'^2 = 
m_3^2 + {1 \over 2} m'^2, \\ 
\Delta m_{12}^2 &\simeq& {2 s^2 c m'^3 \over m_3} = {m'^3 \over \sqrt 2 m_3}.
\end{eqnarray}
Using $m_3 = 0.05$ eV, $m' = 0.016$ eV, we find $\Delta m^2_{atm} \simeq 
2.6 \times 10^{-3}$ eV$^2$, and $\Delta m_{sol}^2 \simeq 5.8 \times 10^{-5}$ 
eV$^2$, in good agreement with data \cite{twelve}.  Note also that $|U_{e3}| 
\simeq 0.22$ (close to the maximum value allowed) in this model, due to the 
form \cite{grla} of Eq.~(8).

Referring back to Fig.~2, we calculate the $\nu_e \nu_\tau$ mass term to be
\begin{equation}
m' \simeq {h \lambda m_\tau^2 v_1 \over 16 \pi^2 v_0 m_\zeta^2}.
\end{equation}
Let $v_0 = v_1$, $m_\zeta = 1$ TeV, then $m' = 0.016$ eV implies 
$h \lambda = 0.8$ MeV.  This is consistent with our assumption that the 
term in Eq.~(7) breaks the assumed $Z_2$ discrete symmetry softly, so 
that $\lambda$ may be naturally small \cite{thooft}.  Note also that $m_\zeta$ 
is assumed to be heavy in order that it does not contribute significantly to 
$\tau \to e \nu_\tau \bar \nu_e$.

To obtain $v_1 = v_2$, we assume that the Higgs potential containing $\Phi$, 
$\eta_{1,2}$ and $\zeta$ is invariant under the interchange of $\eta_1$ and 
$\eta_2$.  In that case, the components of $(\eta_1 - \eta_2)/\sqrt 2$ are 
mass eigenstates.  If they are the lightest scalars, they would be stable 
because neither $\eta_1$ nor $\eta_2$ could decay into light fermions [see 
Eq.~(4)].  However, the $\eta_1 - \eta_2$ interchange symmetry cannot be 
exact because of Eq.~(4) and other terms of the Standard Model; hence we 
expect some small mixing between $(\eta_1 - \eta_2)/\sqrt 2$ and $\Phi$, 
which will allow it to decay, but with an enhanced lifetime.  Note also 
that $X \to -X$ under the interchange of $\eta_1$ and $\eta_2$; hence 
even (odd) states under this symmetry may decay into lighter odd (even) 
states + $X$ (either real or virtual) in this model.

Assuming the typical range of $M_X/g_X \sim 200$ GeV for explaining the muon 
anomalous magnetic moment, one expects interesting phenomenological 
signatures of the $X$ boson at present and future high-energy collider 
experiments. Let us discuss them one by one.

Firstly one can search for the $Z \rightarrow \bar ffX$ decay in the
LEP-I data, where $f = \mu,\tau$ or $\nu_{\mu,\tau}$.  The squared
decay amplitude averaged over the $Z$ polarizations is 
\begin{eqnarray}
|\bar M|^2 &=& 16 E_1 E_2 g^2_X g^2_Z \Bigg[(1+\cos\theta) \left\{{1
\over (M_Z - 2E_1)^2} + {1 \over (M_Z - 2E_2)^2}\right\} \nonumber
\\[2mm] && + {4(1-\cos\theta) \over (M_Z-2E_1)(M_Z-2E_2)} \left\{1 -
{E_1 + E_2 \over M_Z} + {E_1 E_2 (1 - \cos\theta) \over
M^2_Z}\right\}\Bigg] 
\label{sixteen}
\end{eqnarray}
where $g^2_Z = (2e^2 / \sin^2 2\theta_W) (I^2_{3f} +
2\sin^4\theta_W Q^2_f - 2\sin^2 \theta_W I_{3f} Q_f)$ and $E_{1,2}$
are the energies of $\bar f,f$ and $\theta$ the angle between them in
the $Z$ rest frame.  In particular the decay $Z \rightarrow \mu^+
\mu^- X$, follows by $X \rightarrow \mu^+ \mu^- \ (BR = 1/3)$, leads
to a clean 4-muon final state.  We have computed this signal
cross-section incorporating a $p_T > 3$ GeV cut on each muon, as
required for muon identification at LEP, and made a comparison with
the ALEPH data \cite{fourteen}.  This corresponds to 1.6 million
hadronic $Z$ events and shows 20 4-muon events against the SM
prediction of 20.0 $\pm$ 0.6.  Moreover the smaller $\mu^+ \mu^-$
invariant mass for all these events as well as the SM prediction is $<
20$ GeV.  Thus the 95\% CL upper bound on the number of signal events
for $M_X > 20$ GeV is 3, corresponding to the 0 observed events.
Fig. 3 shows the resulting lower limit on $M_X$ as a function of $g_X$,
i.e. $M_X > 50 (70)$ GeV for $g_X \gsim 0.1 (1)$.  

Secondly the
model predicts a small deviation from the universality of $Z$ boson
coupling to $e^+e^-$, $\mu^+\mu^-$ and $\tau^+\tau^-$ channels, since
the latter ones have an extra one-loop radiative correction from $X$.  
The resulting contribution to the $Z$ width is
given by \cite{fifteen} 
\begin{equation}
{\Delta \Gamma \over \Gamma} = - {g^2_X \over 4\pi^2} \left[{7\over4}
+ \delta + \left(\delta + {3\over2}\right) \ell n \delta + (1 +
\delta^2) \left\{Li_2 \left({\delta \over 1+\delta}\right) + {1\over2}
\ell n^2 \left({\delta \over 1+\delta}\right) - {\pi^2 \over
6}\right\}\right], 
\label{seventeen}
\end{equation}
where $\delta \equiv M^2_X/M^2_Z$, and $Li_2 (x) = - \int^x_0 (dt/t)
\ell n(1 - t)$ is the Spence function.  The measured $Z$ partial widths at
LEP-I \cite{sixteen},
\begin{equation}
\Gamma_e = 84.02 \pm 0.14 \ {\rm MeV}, \ \Gamma_\mu = 84.00 \pm 0.21 \
{\rm MeV},
\label{eighteen}
\end{equation}
correspond to a 95\% CL limit of $\Delta \Gamma/\Gamma < 0.006$ on
adding the two errors in quadarature.  The resulting upper limit on
$g_X$ is shown in Fig. 3 as a function of $M_X$.  It does not give any
serious constraint on the mass or coupling of the $X$ boson.

We have estimated the signal cross-section for $X$ boson production at
LEP 200 and LC energies via $e^+e^- \rightarrow \mu^+ \mu^- X$,
followed by the $X \rightarrow \mu^+ \mu^-$ decay.  The squared
Feynman amplitude for $e^+e^- \rightarrow \mu^+ \mu^- X$ was evaluated
using the FORM program \cite{seventeen}.  The resulting 4-muon signal
cross-sections are shown in Fig. 4 for $g_X = 1$, where we have again
imposed a $p^\mu_T > 3$ GeV cut as required for muon identification.
The signal can be easily distinguished from the SM background of 
Drell-Yan pairs via the clustering of a $\mu^+\mu^-$ invariant mass at
$M_X$.  Thus a signal size of $\sim 10$ events should be adequate for
discovery of the $X$ boson.  With the integrated luminosity of $\sim
0.7 \ {\rm fb}^{-1}$ at LEP 200, this corresponds to a signal
cross-section of $\sim 10$ fb.  Thus we see from Fig. 4 that the LEP
200 limit on $X$ mass is $M_X > 60$ GeV for $g_X = 1$, which is no
better than the LEP-I limit.  With the projected luminosity of $\sim
100 \ {\rm fb}^{-1}$ at LC, a signal cross-section of 0.1 fb should be
viable.  This corresponds to a discovery limit of $M_X = 300 (500)$
GeV at LC 500 (LC 1000) for $g_X = 1$.  Note that the signal
cross-section scales like $g^2_X$.  Thus the LC discovery limit goes
down to 200 (250) GeV for $g_X = 0.3$ and to 100 GeV for $g_X = 0.1$.

We have also estimated the $X$ signal cross-section for TEV 2 and LHC
energies of 2 and 14 TeV respectively.  In each case we have computed
the 3-muon and 4-muon signals from $u\bar d \rightarrow \mu \nu X$ and
$u\bar u (d\bar d) \rightarrow \mu^+ \mu^- X$ respectively, followed
by $X \rightarrow \mu^+\mu^-$.  We have imposed a $p^\mu_T > 10$ GeV
and $|\eta_\mu| < 2.5$ cut on each muon as required for muon
identification at these colliders.  The resulting signal
cross-sections are shown in Fig. 5.  Even in this case we expect that
the clean 3-muon and 4-muon signal events can be distinguished from
the SM background via the clustering of a $\mu^+\mu^-$ invariant mass
at $M_X$.  Thus we again consider a signal size of $\sim 10$ events as
adequate for the discovery of $X$ boson.  With the expected luminosity
of $\sim 2 ~{\rm fb}^{-1}$ in Run II of the Tevatron, this corresponds
to a signal cross-section of $\sim 10$ fb.  This means a discovery
limit of $M_X = 70$ GeV for $g_X = 1$, i.e. similar to LEP 200.  The
projected luminosity of 100 fb$^{-1}$ at LHC implies a viable signal
cross-section of 0.1 fb.  This corresponds to a discovery limit of 400
GeV for $g_X = 1$, going down to 200 (100) GeV for $g_X = 0.3 (0.1)$.
These are very similar to the corresponding discovery limits of LC.
While they do not exhaust the full range of $M_X/g_X$, they do cover
the interesting range of $M_X/g_X \sim 200$ GeV.  Finally one expects
copious production of the $X$ boson at muon colliders right upto $M_X
= \sqrt{s}$, because of its gauge coupling to the $\mu^+ \mu^-$ pair. 

In conclusion, we have proposed in the above a verifiable explanation of 
the possible discrepancy of the newly measured muon anomalous magnetic 
moment as coming from the realization of the gauged $L_\mu - L_\tau$ 
symmetry at the electroweak energy scale.  Our specific model has the 
added advantage of allowing a simple neutrino mass matrix which can 
explain the present data on atmospheric and solar neutrino oscillations. 
We discuss the phenomenology of the associated gauge boson $X$ and show 
that it can indeed be relatively light, i.e. $M_X/g_X \sim 200$ GeV, and 
be observed through its distinctive decay into $\mu^+ \mu^-$ at future 
high-energy colliders.

We are grateful to Rajeev Bhalerao, Utpal Chattopadhyay and Dilip
Kumar Ghosh for computing advice.  The work of EM was supported in part by the
U.~S.~Department of Energy under Grant No.~DE-FG03-94ER40837.  

\bibliographystyle{unsrt}


\newpage

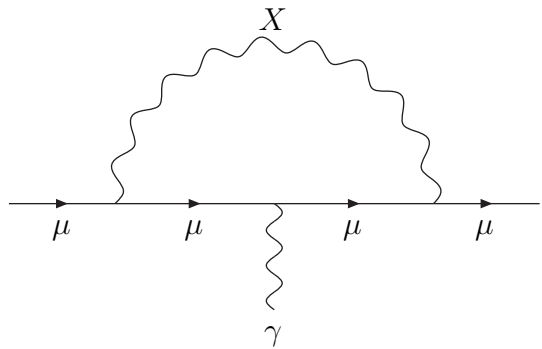
\begin{figure}
\begin{center}
\begin{picture} (270,150) (0,0)
\ArrowLine(35,50)(75,50)
\ArrowLine(75,50)(135,50)
\ArrowLine(135,50)(195,50)
\ArrowLine(195,50)(235,50)
\Photon(135,50)(135,10){3}{3}
\PhotonArc(135,50)(60,0,180){3}{10}
\Text(55,40)[]{$\mu$}
\Text(105,40)[]{$\mu$}
\Text(165,40)[]{$\mu$}
\Text(215,40)[]{$\mu$}
\Text(135,0)[]{$\gamma$}
\Text(135,120)[]{$X$}
\end{picture}
\end{center}
\caption{Contribution of $X$ to muon magnetic moment.}
\end{figure}

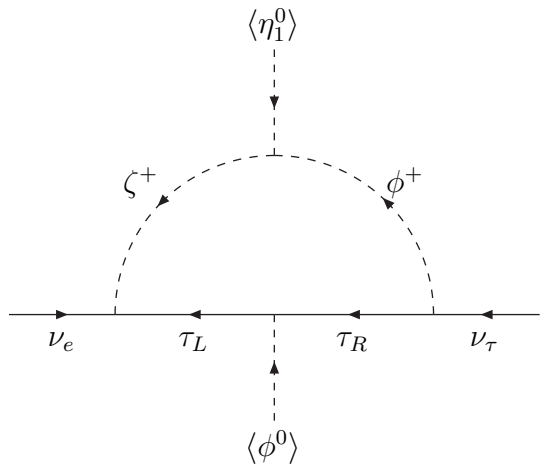
\begin{figure}
\begin{center}
\begin{picture}(300,150)(0,0)
\ArrowLine(45,40)(85,40)
\ArrowLine(145,40)(85,40)
\ArrowLine(205,40)(145,40)
\ArrowLine(245,40)(205,40)
\DashArrowLine(145,0)(145,40){3}
\DashArrowArc(145,40)(60,90,180){3}
\DashArrowArc(145,40)(60,0,90){3}
\DashArrowLine(145,140)(145,100){3}
\Text(65,30)[]{$\nu_e$}
\Text(115,30)[]{$\tau_L$}
\Text(175,30)[]{$\tau_R$}
\Text(225,30)[]{$\nu_\tau$}
\Text(145,150)[]{$\langle \eta_1^0 \rangle$}
\Text(95,90)[]{$\zeta^+$}
\Text(195,90)[]{$\phi^+$}
\Text(145,-10)[]{$\langle \phi^0 \rangle$}
\end{picture}
\end{center}
\caption{Radiative contribution to the $\nu_e \nu_\tau$ mass.}
\end{figure}

\newpage

\begin{figure}
\begin{center}
\vspace*{3.5in}
\includegraphics{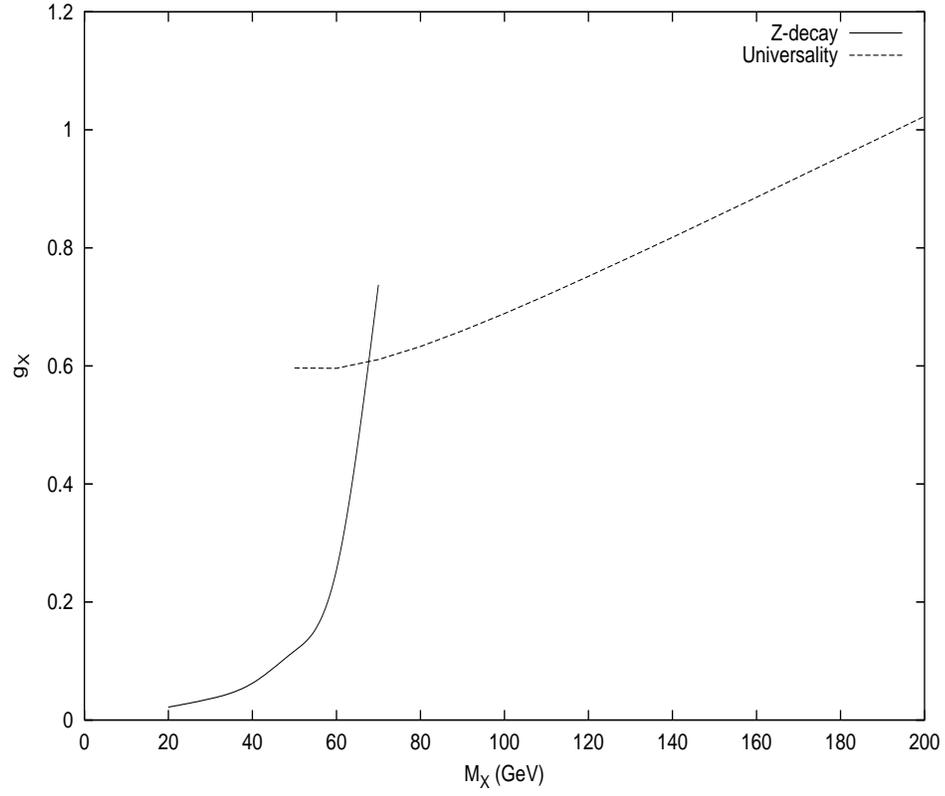}
\label{fig:0139_fig3}
\end{center}
\caption{The LEP-I constraints on the mass and coupling of the $X$
boson from $Z \rightarrow X \mu\mu$ decay and the universality of
$Z$ coupling to the $e^+e^-$ and $\mu^+\mu^-$ channels.  The region
above the curves is excluded at 95\% CL.}
\end{figure}

\newpage

\begin{figure}
\begin{center}
\vspace*{3.5in}
\includegraphics{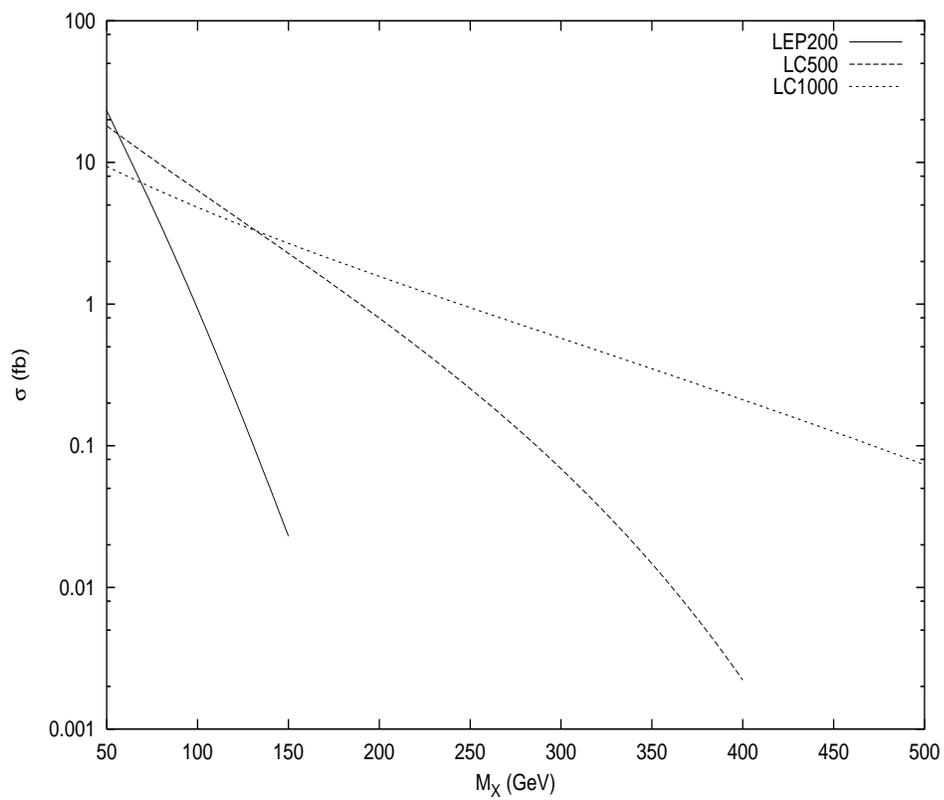}
\label{fig:0139_fig4}
\end{center}
\caption{The $X$ boson signal cross-section in the 4-muon channel at
LEP and LC energies of 200, 500 and 1000 GeV for $g_X = 1$.}
\end{figure}

\newpage

\begin{figure}
\begin{center}
\vspace*{3.5in}
\includegraphics{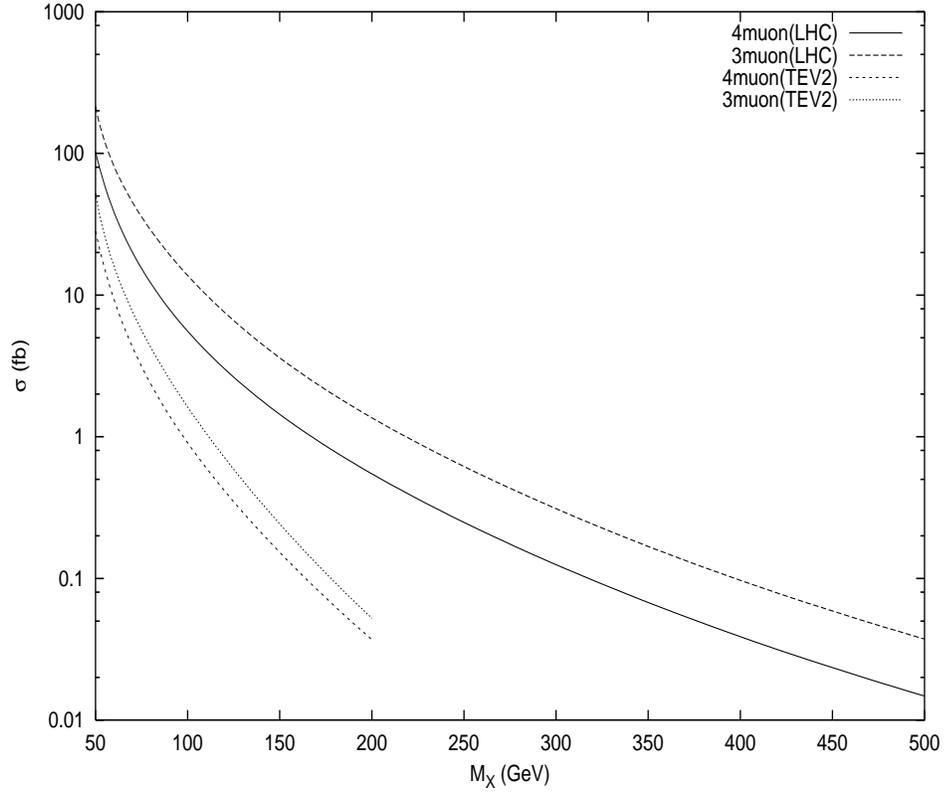}
\label{fig:0139_fig5}
\end{center}
\caption{The $X$ boson signal cross-section in the 3-muon and 4-muons
channels at the Tevatron and LHC energies for $g_X = 1$.}
\end{figure}

\end{document}